# A Multiagent CyberBattleSim for RL Cyber Operation Agents

Thomas Kunz[1], Christian Fisher, James La Novara-Gsell, Christopher Nguyen
Systems and Computer Engineering
Carleton University
Ottawa, Canada
tkunz@sce.carleton.ca, {christianfisher, jameslanovaragsell, chrisnguyen3}@cmail.carleton.ca

Li Li
Defence Research and Development Canada
Ottawa, Canada
li.li@drdc-rddc.gc.ca

*Abstract*—Hardening cyber physical assets is both crucial and labor-intensive. Recently, Machine Learning (ML) in general and Reinforcement Learning RL) more specifically has shown great promise to automate tasks that otherwise would require significant human insight/intelligence. The development of autonomous RL agents requires a suitable training environment that allows us to quickly evaluate various alternatives, in particular how to arrange training scenarios that pit attackers and defenders against each other. CyberBattleSim is a training environment that supports the training of red agents, i.e., attackers. We added the capability to train blue agents, i.e., defenders. The paper describes our changes and reports on the results we obtained when training blue agents, either in isolation or jointly with red agents. Our results show that training a blue agent does lead to stronger defenses against attacks. In particular, training a blue agent jointly with a red agent increases the blue agent's capability to thwart sophisticated red agents.

## I. INTRODUCTION

The modern world relies heavily on the correct operation of communication networks in general and the Internet in particular. Rogue actors regularly attempt to disrupt such networks, comprising confidentiality, availability, and/or the integrity of essential information. Organizations and governments are consequently interested in hardening their systems, both by learning about possible attack vectors, and suitable defenses. This is typically carried out as contests between a red team attacking the network and a blue team defending it. The key challenge that arises is the scalability of the expert blue and red teams. Cyber defence and attack emulation tools provide the blue and red teams with the capability to automate the IT work flow, enabling the staging, scripting commands and filling in payloads. However, the sequential decision making at each action step when using them relies on the human cyber expert.

Alternatively, Machine Learning (ML) in general, and Reinforcement Learning (RL) in particular, hold great promise for the development of autonomous red and blue agents. Depending on the sophistication of the RL training environment, trained agents may exhibit human-level or even superior intelligence in the choice of their actions. It is therefore essential for operators of networks and government agencies tasked with defending a nation's cyber infrastructure to explore these capabilities and to build appropriate defenses.

At the RL core is a training environment that both mimics the real world and allows agents to be trained. Recently, for autonomous cyber operations, a number of such training environments have been proposed and described in the literature, including CybORG [16], CyGIL [9], or CyberBattleSim [10] and are reviewed below. CyberBattleSim supports training red agents (i.e., attackers), and we extended it to support training blue agents (i.e., defenders) as well. The paper describes the changes we made and reports on the results we obtained when training blue agents in the canonical "capture-the-flag" game. Our extension also allows the blue and the red agents to be trained jointly. Using these enhanced capabilities, we can then draw insights and lessons as to how to arrange for appropriate training scenarios to ensure strong autonomous agents, in particular for the blue agent.

The paper is organized as follows. Section II reviews related work on hardening the network and computer components of cyberphysical systems. It focuses particularly on existing training environments for RL-based agents. Section III introduces CyberBattleSim and the changes we made to enable blue agent training, as well as joint agent training. Section IV reports on our experimental results, and Section V concludes the paper with a discussion of ongoing and future work.

## II. RELATED WORK

### A. Non-ML Approaches

Traditionally, testing a network and hardening its defenses is a very labor-intensive process, with human teams playing the role of red and blue agents. Such games require lengthy preparation and setup phases, and involve highly skilled/trained personnel, see for example the description of RMC's CyberX initiative [6], and prior to that the U.S. National Security Agency's Cyber Defence Exercise [14].

Ultimately, since human ingenuity is required both for success attacks and successful defenses, such exercises, expensive and cumbersome as they are to organize, will always have a place. However, many exploits and weaknesses exist in deployed systems and networks that can be targeted using automated means. Consequently, a number of tools have been developed to automate penetration testing [1] of networks and systems, based on known weaknesses (see for example the

---

[1] Corresponding Author

Metasploit Framework [4]). Such tools can be utilized by the red team to pick "low-hanging" fruits, automating attacks on poorly defended network infrastructure. Alternatively, blue teams can use this tool to ensure that their system is hardened at least against known, common exploits. However, at the end of the day, such tools display little intelligence in how they attack a network. They also are limited in that they use (and exploit) known vulnerabilities and attack vectors only. It is also non-trivial with these tools to experiment with and evaluate defense strategies that are focused on deceptions.

*B. ML Approaches*

More recently, Machine Learning has shown great promise to automate tasks that otherwise would require significant human insight/intelligence: image recognition, language translation, etc. Consequently, researchers have started to explore the capability of Machine Learning (ML) in the realm of cyber security [15]: can we use ML approaches to either improve the capabilities of the attacker, and/or we can strengthen the network defenses. In our work, we are exploring the use of ML to develop autonomous agents that can replace (or complement) the red and blue teams, called red and blue agents respectively. Of particular interest within ML in the context of Cyber Operations has been Reinforcement Learning (RL) [11]: agents take actions (either offensive or defensive) based on their observation of the state of the network. These actions lead to changes in the state and the agent receives a reward (both positive and negative). For example, actions that allow a red agent to learn about additional devices in the network or the existence of specific services on a known device will be rewarded positively. Network activities that do not yield new information but run the risk of detection by a blue agent will be rewarded negatively (i.e., penalized). For a blue agent, possible actions may be the reimaging of systems or the deployment of decoys such as honeypots or honeytokens [2, 18]. If these actions result in the eviction of an attacker on a host, or successfully fool attackers, they will be positively rewarded. On the other hand, spending resources on actions that do not deter or deceive an attacker will be penalized (negatively rewarded). RL uses this basic mechanism to train, over many iterations, a specific sequence of actions, with the goal of maximizing an agent's long-term overall reward. For example, a trained red agent tries to compromise a system or expose confidential data stored in a specific server, a trained blue agent aims, in the long run, to ensure continuous operation of a network and the confidentiality of its essential data.

Common to all RL-based approaches is the need for a training environment. CyGIL [9] is an emulated RL/DRL training environment for cyber operations. It allows to train a red agent in a realistic environment, using scans and exploits that exist in state-of-the-art attack frameworks. It provides an interface to OpenAI Gym that allows the red agent access to the RL training algorithms available through this framework. While CyGIL strives for realism, it is limited in scope, as training in the emulated environment takes significant amounts of time. As reported by the authors, a single training run for a relatively small network scenarios took multiple days to complete.

CyberBattleSim [10] investigates how autonomous agents operate in a simulated enterprise environment using high-level abstraction of computer networks and cybersecurity concepts. The original release focused on training a red agent only. Similar to CyGIL CyberBattleSim provides an interface to OpenAI Gym to train agents via available RL algorithms.

CybORG [16] provides an experimental environment for training and development of cyber human and autonomous agents. It contains a common interface for both an emulated network environment that uses cloud based virtual machines, and a simulated network environment. The explicit goal of its developers is to provide a high-fidelity simulation environment that facilitates the rapid training of autonomous agents that can then be tested on real-world systems. Similar to both CyGIL and CyberBattleSim, CybORG provides an interface to OpenAI Gym, and both red and blue agents can be trained with RL algorithms found in that framework. The paper shows, via small network of three hosts in a line topology, that agents can be trained in the simulation environment and then execute successfully in the virtual environment using professional security tools. Yet the network was small and there was a non-trivial number of failures to transfer trained agents from simulation to emulation (about 1/3 of all cases).

CyGIL and CybORG explicitly address the challenge of training agents in a realistic environment, ensuring that a trained agent can successfully be deployed in a real network. Actions are taken from state-of-the-art attack tools. CyberBattleSim is less concerned with the transferability of the trained agent into a real deployment. Rather, it allows researchers to study alternative deployments and attack/defense strategies, evaluating their respective pros and cons. Other researchers have expanded CyberBattleSim to study, at this conceptual level, various defensive strategies such as the costs and benefits of deploying honeypots and honeytokens in [18]. We are interested in deriving insights into how to arrange for agent training, necessitating a solution that allows us to extensively and exhaustively test different arrangements. We therefore selected CyberBattleSim as the basis for our work, expanding its capabilities by interfacing it more closely with a popular library of RL algorithms, allowing the training of blue agents, and joint training of red and blue agents.

III. CYBERBATTLESIM

This section first describes the original CyberBattleSim. We will then describe our changes to enhancing CyberBattleSim, addressing some existing shortcomings, adding a trainable defender, and enabling multi-agent training, resulting in a solution we call MARLon (Multi-Agent Training ON CyberBattleSim).

*A. Original CyberBattleSim*

CyberBattleSim is a Python OpenAI Gym environment to model a network security problem. The environment CyberBattleSim creates is a network of nodes connected by a series of different network connection protocols. An example of such a network is shown in Fig. 1. This network has a number of nodes, such as a Client, Website, GitHubProject and AzureStorage, which are all connected by their required protocols. CyberBattleSim models network security problems by introducing an attacker into the network The goal of the attacker is to take ownership of a portion of the network by exploiting vulnerabilities in the network while the defender of the network, attempts to detect the attacks and mitigate the impact on the system by evicting the attacker. In this snapshot, nodes coloured in red have been compromised already. CyberBattleSim provides a trainable attacker which learns based on allotted rewards determined by the outcome of its selected action from the provided action space. If the attacker performs an action which is not possible in the space, it loses rewards. Otherwise, the attacker performed a valid action resulting in rewards based on the effectiveness of the action.

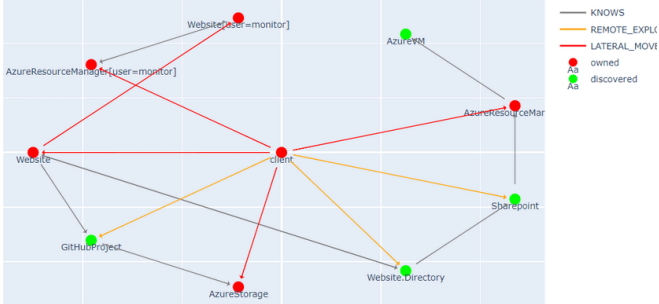

Figure 1: Example Network Topology in CyberBattleSim

The attacker has three main action types, which are the local vulnerability, remote vulnerability and connect actions. The local vulnerability exploits an existing vulnerability on a node to obtain credentials. The remote vulnerability exploits a vulnerability on another node to obtain credentials. Finally, the connect action completes a lateral movement by using saved credentials to infect and take over a remote node. Each of these three types of actions require additional information for execution, such as the node to act on, the port on which to connect to, or the credentials to use. This additional information is included as individual fields in the action space for the agent to use. As a result, the attacker action space grows quickly during command parameterization under each action type.

The observation space of the CyberBattleSim attacker consists of a number of fields, grouped in two sections. The first section contains information gained from the last step, and the second section contains information gained over the episode. Some examples from the first section are the count of newly discovered nodes, whether a lateral move was successful, or newly discovered credentials. Examples of the second section are the total number of discovered nodes, discovered node properties, or all known credentials. All of these fields combine to describe the status of the network as the attacker views it.

The reward calculation for the attacker after each step of action is determined by the action type, the result of the action, and the value of the node being acted upon as defined by the CyberBattleSim environment.

Finally, in addition to the trainable attacker, CyberBattleSim also includes a non-trainable defender to play against the attacker during its training. The defender works on a static probability model. The built-in defender's only available action is to randomly choose a node to reimage, and the success of this action is determined by a configured probability. The defender is also constrained to ensure a certain level of network availability at all times. Thus the defender cannot lock out the attacker by reimaging all nodes at once. The defender is not an agent, but is built directly into the CyberBattleSim environment.

To train or evaluate an agent, the network topology must first be selected. There are a few built in CyberBattleSim topologies, such as the Chain Network or the ToyCTF Network. The ToyCTF network is used in our experiment as it is more complex. As shown in Fig. 1, it consists of 10 nodes, each subject to specific vulnerabilities. The goal of the attacker is to take ownership of critical nodes in the graph (e.g., Azure and Sharepoint resources). Connectivity among nodes is constrained by firewall rules. For example, the website node allows incoming connections from any node, and has three vulnerabilities that allow an attacker to detect incoming connections (exposing other hosts), find a browseable directory with a textfile, and to obtain additional credentials by reading this file. When training starts, the attacker agent, and optionally, the (random) defender will begin taking steps and acting on the network. In each step, the attacker chooses an action, then has the CyberBattleSim environment execute this action. Within this same step, the defender will have an opportunity to take its own action as well. The episode will terminate after a number of steps, and the cumulative reward attained by a trained attacker is used to compare various training approaches. As described in the CyberBattleSim documentation, for episode lengths of 1,500 steps, cumulative rewards range from around 270 for a trained Q agent to up to 430 for a trained Deep Q agent.

*B. Enhancements to Environment Interface*

To create the multi-agent environment in CyberBattleSim, the defender agent support must follow the RL model which includes first the action and observation spaces for the agent. CyberBattleSim however has a complicated structure in providing its action space and cannot handle invalid actions. The agent in training must not issue any invalid action or the CyberBattleSim environment will crash. As most of the actions in the action space can be invalid, this limitation forces the agent in training to be "action syntax aware" so that it launches only valid actions to the environment. Consequently, CyberBattleSim prevents the agent training from using directly many of the state-of-the-art RL/DRL algorithm libraries which are action space agnostic. In addition, the structure of the observation space in CyberBattleSim also is found to have certain compatibility issues when tested using selected RL/DRL libraries, e.g., the `stable-baselines3` [12] library.

To fix the interface issues, we implemented first a new class called `AttackerWrapper` on top of CyberBattleSim to handle invalid actions so that the agent is freed from being action syntax aware. This feature also brings extendibility to CyberBattleSim in facilitating new additional actions in the future. The wrapper conforms to the open AI Gym interface to correct also the observation space incompatibility issues. Additionally, the wrapper can apply an invalid action penalty to the reward function to discourage agents from selecting invalid actions.

*C. Trainable Defender*

For the defender agent training support, four main features are implemented in a class called `DefenderWrapper`: the defender's observation space, action space, reward function, and an action validity checker. With new features added in the wrapper, the underlying CyberBattleSim environment is leveraged without having to change any of its code.

The defender's action space consists of five basic actions. These actions existed in CyberBattleSim previously, but were not utilised by the built-in defender. The defender wrapper collects them and creates the defender's action space:
- Reimage Node: Reset a node, removing any attacker control
- Block Traffic: Use a firewall rule to block a specific type of traffic on a node
- Allow Traffic: Allow a specific type of traffic on a node
- Stop Service: Stop a running service on a node
- Start Service: Start a stopped service on a node

These five actions all require parameters to be selected for execution. For example, if the defender agent chooses the "reimage node" action, the agent must also choose which node to reimage. The parameter selection can lead to a valid or an invalid action. For example, if the agent chooses to reimage node seven and this node is currently offline, this action would be invalid.

The defender's observation space consists of four fields. These fields were chosen because each one relates to a specific action in the defender's action space.
- A list of all infected vs non-infected nodes in the network
- The status of all incoming firewall rules in the network
- The status of all outgoing firewall rules in the network
- The status of all services in the network

The defender wrapper validates any action taken to ensure it is valid before passing it to the underlying CyberBattleSim environment. If the action is found to be invalid, the defender will do nothing and return an invalid action penalty as reward.

Unlike the attacker, which uses the built-in CyberBattleSim reward calculation, there is no existing support for determining how "good" a defender action is. A new function is required for the defender. We decided on the following approach: the defender's reward is calculated by negating the attacker's last reward. Furthermore, if the defender breaks its network availability constraint, it will lose the game, and a large negative reward will be applied. These choices were made because the defender's purpose is to minimise the attacker's progress and therefore reward. This means the defender's maximum reward is zero, and will become negative for any progress the attacker makes. This does however directly link the attacker's learning progress to the defender. If the attacker makes little to no progress due to its poor performance, the defender will receive a "high" reward (close to zero) for nothing. Fig. 2 shows an overview of how our system, MARLon, works in conjunction with the CyberBattleSim environment. The attacker agent passes the action to the attacker wrapper, which in turn will pass the validated action to the environment. The environment will generate the reward and observation, which will be passed to the attacker wrapper, and then back to the attacker agent. The defender agent similarly selects an action to take, which will be passed to the defender wrapper. The wrapper will validate the action, then forward it to the same environment. The wrapper will then generate its own observation and reward for the defender, and pass those back to the defender agent.

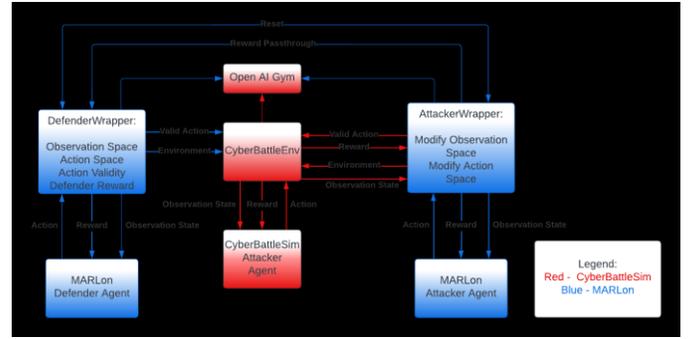

Figure 2: MARLon high level diagram.

*D. Multi-Agent Learning Algorithms*

In order to train both attacker and defender agents at the same time, we started from the `stable-baseline3` 'learn' algorithm and modified it so it works with two agents at once by allowing them to take turns performing actions and receiving rewards. Both attacker and defender agents will indirectly operate on the same CyberBattleSim environment through their respective wrappers. In the current experiment, the `stable-baseline3` library is used for all agent training algorithms.

A consequence of training on two wrappers at the same time is that when an episode ends, and reset is called, it needs to be coordinated with the other wrapper. Whichever wrapper receives a reset call first should forward that call to the underlying CyberBattleSim environment and additionally notify the other wrapper that a reset has occurred. The other wrapper must not reset the CyberBattleSim environment since it has already been reset, yet must appear to reset to the baseline agent it is paired with so that both agents are aware a new episode is starting, which is important during training.

IV. RESULTS

We train red and blue agents using the PPO and A2C algorithms from the `stable-baselines3` distribution. For convenience, we refer to an agent trained with the PPO algorithm as, for example, the PPO Red Agent. If the PPO training happened jointly with training the opposite agent, we add the term MARL, so for example PPO MARL Blue Agent. When agents are jointly trained, we will use the same training algorithms on both sides.

The agent models are trained to maximise their accumulated rewards per episode for a total of 300,000 timesteps. Each episodes consists of at most 2,000 timesteps and we train and evaluate all agents in the ToyCTF CyberBattleSim environment discussed earlier. The blue agent is constrained to maintain 60% network availability. If that constraint is broken, the blue agent loses, receives a 5000-reward penalty, and the episode is stopped prematurely.

Training for at least 300,000 timesteps is required for blue agents to consistently learn how to avoid breaking their availability constraint. Based on the range of values CyberBattleSim uses in benchmarks and our initial experiments, 2,000 steps for the maximum episode length are found to be the best balance for multi-agent learning. The red agent would have not sufficient time to make progress if we selected fewer steps. More steps per episode did not lead to a noticeable improvement in the performance of trained agents. Once an agent training is completed, we save the model parameters in a separate file. We also used the original CyberBattleSim distribution to separately train Tabular Q and Deep Q red agents in the same scenario, saving the trained models for future evaluation as well.

For evaluation purposes, we instantiate a trained agent from the stored model parameter file. We then again run multiple episodes of at most 2,000 steps (typically 25 episodes). However, the simulation performs no training but simply allows us to evaluate any combination of agents against each other. At the end of each episode we collect the accumulated rewards for each agent and report the average value, and, where appropriate, 95% confidence intervals.

## A. Effects of Invalid Actions on Training Performance

As discussed earlier, agents frequently select invalid actions, which are then replaced by our wrappers with valid, randomly selected actions. In a first scenario, we tested the effects of various ways to handle invalid actions via the reward function returned by our wrappers during the training of an A2C red agent against an undefended network.

TABLE 1: A2C EVALUATIONS WITH VARIOUS INVALID ACTION REWARDS

| Red Agents | Score |
|---|---|
| A2C - Invalid action penalty (reward of -1) | 0 |
| A2C - No invalid action penalty | 283 |
| A2C - 0 invalid action reward | 359 |

Table 1 shows the results of evaluations of several types of invalid action rewards. In the first example, we use negative reinforcement by applying a small penalty every time the action is invalid. In this example the red agent decided to only choose actions that it knew were always valid to avoid the penalties, even if they returned no positive reward. The red agent never chose the 'connect' action, as many parameter combinations for this action will lead to an invalid action and hence be rewarded negatively. The red agent consequently failed to gain rewards from moving through the network. The agent effectively becomes paralyzed by the fear of the being negatively rewarded when choosing invalid actions.

In the second example, when an invalid action is selected and redirected to a random valid action, we will not modify the reward of the redirected action. In this example, the red agent achieves much better performance, obtaining a cumulative reward of 283 reward. It struggled to do better since it becomes difficult for the red agent to distinguish between valid and invalid actions. When we redirect invalid actions to random valid actions, we do so without the agent's knowledge, so it may be misled as to which actions yield high rewards.

Finally, the last example is what can be called positive reinforcement. Instead of penalising the reward of an invalid action, we will explicitly assign a reward of 0, independent of the reward that the randomly selected valid action achieved. This should encourage the red agent to try new actions since no reward is gained or lost from invalid actions. This means the red agent has nothing to lose and everything to gain by trying to find new valid actions that yield rewards. This approach resulted in another significant jump in cumulative reward to 359, proving to be the most effective of the three alternatives. Going forward, all agents were therefore trained with this approach.

## B. Impact of Network Availability Requirement

One game requirement, as discussed above, is the need for the blue agent to maintain a specific network availability level. When the network availability drops below the threshold (60%), the blue agent is assigned a large negative reward and the training episode terminates prematurely. Over time, the blue agent is expected to learn to avoid this. In a second set of experiments we therefore tested the effect of this requirement on the red agent's performance. These experiments were done using a combination of basic red or blue agent, and a PPO trained red agent. A basic agent selects valid actions purely at random, meaning it does not learn. The PPO Red Agent was trained in a network with no blue agent. The results of the evaluation runs, averaged over 25 repetitions, are displayed below.

TABLE 2: SINGLE TRAINED PPO RED AGENT VS BASIC BLUE AGENT

| Red Agent | Blue Agent | Episode Length | Red Score | Blue Score |
|---|---|---|---|---|
| Basic | Basic | 140 | 61.2 | -5061.2 |
| PPO | None | 2000 | 320.2 | |
| PPO | Basic | 140 | 72.6 | -5072.6 |

The basic versus basic scenario provides a baseline for how random red and blue agents perform. Since the blue agent is taking random actions, it will quickly break the network availability constraint and loses, causing the short episode length of 140. The cumulative red agent reward is relatively low since the episode ends quickly. If a PPO trained red agent attacks a network without a blue agent, the average episode length will be 2,000 timesteps, with an average cumulative reward of 320. However, if we add a random blue agent to the PPO Red Agent, the episode length drops to 140 again, with the red agent's cumulative reward dropping to approximately 72. The random blue agent, as before, caused the episode to terminate prematurely due to the violation of the network availability requirement. The red agent, being trained, still does not have a lot of opportunity to accumulate rewards, leading to only a slightly better performance than the random red agent.

## C. Multi-Agent Reinforcement Learning With Network Availability Constraints

The blue agent causing premature resets in the environment (due to violating the network availability constraint) poses a much larger problem in a multi-agent training scenario. We trained a PPO Red Agent and a PPO Blue Agent simultaneously. These agents compete during training to gain control of the network. The results of those agents are shown below, again after training the agents and averaging over 25 evaluation runs. In the first row, the average episode length is 2,000, but results in an average cumulative reward of 6 for the red agent, and -6 for the blue agent. This seems to imply a very positive outcome for the blue agent: In 2,000 timesteps the red agent achieved almost nothing (low accumulated reward). If the same PPO Red

Agent is evaluated against no blue agent, also seen in Table 3, it is not doing any better (and, for example, significantly worse than the basic red agent in Table 2). This indicates that the PPO Red Agent, trained against a PPO Blue Agent, not only failed to learn any promising vector of attack, it in fact learned to be far worse than random.

TABLE 3: PPO MARL EVALUATIONS

| Red Agent | Blue Agent | Episode Length | Red Score | Blue Score |
|---|---|---|---|---|
| PPO MARL | PPO MARL | 2,000 | 6 | -6 |
| PPO MARL | None | 2,000 | 6 | |

The root cause is because, during training, the blue agent, violating network availability, causes frequent training episode terminations. It takes the blue agent a while to learn how to not accidentally lock down the network too much and break its network availability constraint. Every time the blue agent breaks the constraint, the environment resets. When the environment resets, the red agent also gets reset. This means that the red agent, through no fault of its own, is forced to restart its attack in the next episode, never progressing past a certain point in the network. This will continue until the blue agent learns to stop breaking its constraint. Once this happens, the red agent has fallen behind the blue agent in terms of progress, being locked out by the blue agent successfully.

To address this issue, a no reset version of the agents was trained. A blue agent violating the availability constraint would still result in a significant negative reward, but would not terminate the current training episode. This allows the blue agent to learn to not break the network availability constraint, but also allows the red agent to learn how to attack in the meantime. A consequence of this change is all training and evaluation episodes will run for the maximum 2,000 timesteps. Evaluation results s of these agents are shown below.

TABLE 4: PPO MARL No-RESET EVALUATIONS

| Red Agent | Blue Agent | Episode Length | Red Score | Blue Score |
|---|---|---|---|---|
| PPO MARL | PPO MARL | 2000 | 96.4 | -96.4 |
| PPO MARL | None | 2000 | 309.7 | |

As seen in Table 4, the red agent achieves a cumulate reward of 96 when paired against a blue agent jointly trained with it. This is noticeably higher than the previous value of 6 shown in Table 3. Furthermore, the MARL No-Reset red agent does better when attacking an undefended network, as it achieves a cumulate reward of 309, which is again much higher than the 6 observed previously. This mean cumulative reward of 309 is also fairly close to the results gathered in the PPO against no blue agent scenario as shown in Table 2. In fact, based on the 95% confidence intervals, the difference in these two values is not statistically significant.

While the cumulative reward of the red agent when facing a blue agent is still not very high, we believe this is due to the blue agent having an innate advantage due to the overall structure of the network. If the network were to change to be larger and have less critical paths which the red agent must capture to advance, or the blue agent were to not be omniscient in its information gathering, the blue agent's advantage may not be present and the red agent may do much better.

*D. Multiagent Results*

We trained all red agents individually in a network without blue agent. Training a blue agent requires the existence of a red agent, so we trained all blue agents against the basic red agent described earlier. We also trained red and blue agents jointly. Once all our agents were trained, we ran numerous evaluations on combinations of red and blue agent. We highlight two sets of comparisons below to indicate some of our observations to-date.

TABLE 5: A2C RED AGENT evaluation results

| Red Agent | Blue Agent | Red Score | 95% CI |
|---|---|---|---|
| A2C | None | 320.6 | 320.6 ± 34.1 |
| A2C | PPO | 127.1 | 127.1 ± 52.1 |
| A2C | PPO MARL | 98.5 | 98.5 ± 22.6 |

The A2C red agent, trained and evaluated in a network without blue agent, achieves a cumulative reward of almost 321. Evaluating the same A2C red agent in a network with PPO and PPO_MARL blue agents, the red agent cumulative reward drops to, on average, 127 and 99 respectively. The MARL trained blue agent was better at preventing the red agent from accumulating rewards than the single trained blue agent. In both cases, adding a trained blue agent significantly reduces the A2C red agent's capability to impact the network within the 2,000 steps of a single episode. The PPO MARL blue agent performed slightly better than the PPO blue agent while having a significantly lower standard deviation, indicating a more consistent performance across multiple evaluation runs.

TABLE 6: PPO/PPO MARL evaluation Red Agent results

| Red Agent | Blue Agent | Red Score | 95% CI |
|---|---|---|---|
| PPO | None | 320.2 | 320.2 ± 31.5 |
| PPO | PPO MARL | 137.8 | 137.8 ± 51.7 |
| PPO MARL | None | 117.9 | 117.9 ± 11.6 |
| PPO MARL | PPO MARL | 96.4 | 96.4 ± 17.4 |

Using PPO and PPO_MARL red agents against the network with no blue agent, as base cases, we see the red agents average a score of 320 and 118 respectively. When adding blue agents in both cases, the cumulative reward of the red agent reduces: for the PPO red agent it dropped from 320 to 138 and for the PPO MARL red agent it dropped from 128 to 96. Unfortunately, this case shows that our MARL blue agent performed worse against a PPO MARL red agent than against an individually trained PPO red agent. However, these results still demonstrate that adding a trained blue agent significantly hinders the red agent's ability to progress and to accumulate rewards.

Overall, the data we collected to-date implies that having a blue agent jointly trained against a red agent is beneficial,

resulting in less progress/lower cumulative reward for the red agent. A suitable training environment for blue agents should contain either strong, independently trained, red agents. Alternatively, training blue agents should happen concurrently with training red agents. The benefits of such a setup (trained against a strong blue agent or trained jointly with a blue agent) for the red agent are much less obvious from our results and require further analysis.

## V. Conclusions

The paper describes our extensions to CyberBattleSim to allow for blue agent training as well as jointly training red and blue agents. We also report some of the results we obtained with this extended training environment. Collecting such an exhaustive set of results would not be possible in an emulated training environment. As our results show, we have to be careful when setting up appropriate mutual training environments to not disadvantage the red agent when the blue agent quickly looses games during the training phase. There is also value in carefully considering how to "reward" invalid actions chosen by an agent, something that occurs quite frequently in our training runs. We obtained the best results by a form of positive reinforcement, indicating to the agent that the chosen action yielded no reward. This ultimately encourages the agent to explore other actions during the training phase.

Our results show, as we would expect, that there is a clear benefit to having a trained blue agent active in the network, reducing cumulative rewards obtained by various trained red agents. More interestingly, the blue agent benefits from being trained jointly with non-trivial red agents, in many cases surpassing the performance that is obtained with a blue agent trained in isolation. The same does not seem to be true for red agents.

Continuing our work on exploring appropriate training scenarios, we started to collect results when training red and blue agents jointly or against each other (the results here trained red agents only against an undefended network and blue agents against a basic attacker). One observation to-date seems to be that such combinations may result on over-specialized agents: for example, a defender trained against a PPO attacker will demonstrate strong performance against such an attacker, but perform poorly against an A2C attacker. To generalize our trained agents, we are exploring how to reduce the observation space dimensionality without loosing the training performance, see for example [8]. The goal is to both speed up the training process, allowing us to explore larger scenarios, and also help when transferring the learned strategies to similar scenarios [19]. To reduce the dimensionality of the action space, we are looking at ideas such as [17]. The authors propose a hierarchical approach to deal with large discrete action spaces, decomposing the action space into smaller subsets and training individual agents in each action subset. Furthermore, carefully arranging how agents gradually learn more complex tasks would realize the idea of curriculum learning as proposed in [3]. As argued in their paper, training agents gradually, exposing them to more (and more complex) concepts in a meaningful manner, increases not only convergence speed, but also improves an agent's generalizability.

## Appendix

The code described in this paper is available on Github at https://github.com/James-LG/MARLon.